\documentclass[letterpaper, 10 pt, conference]{ieeeconf}  

\IEEEoverridecommandlockouts                              

\usepackage{graphicx}
\usepackage{dirtytalk}
\usepackage{etoolbox}
\usepackage{booktabs}
\usepackage{multirow}

\usepackage[mathscr]{euscript}

\usepackage[table,xcdraw]{xcolor}

\usepackage{hyperref}

\usepackage{amsmath}
\usepackage{amssymb}

\usepackage{algorithm}
\usepackage{algpseudocode}
\usepackage{enumerate}

\usepackage{threeparttable}
\usepackage{scalerel}
\usepackage{tikz}
\usepackage{cite}

\title{\LARGE \bf
A General Pipeline for Glomerulus Whole-Slide Image Segmentation
}

\author{Quan Huu Cap
\thanks{Quan Huu Cap is with the AI Development Department, Aillis, Inc., Tokyo, Japan. Email: {\tt\small quan.cap@aillis.jp}}
}

\begin{document}

\maketitle
\thispagestyle{empty}
\pagestyle{empty}

\begin{abstract}
    Whole-slide images (WSI) glomerulus segmentation is essential for accurately diagnosing kidney diseases. 
In this work, we propose a general and practical pipeline for glomerulus segmentation that effectively enhances both patch-level and WSI-level segmentation tasks. 
Our approach leverages stitching on overlapping patches, increasing the detection coverage, especially when glomeruli are located near patch image borders. 
In addition, we conduct comprehensive evaluations from different segmentation models across two large and diverse datasets with over 30K glomerulus annotations. 
Experimental results demonstrate that models using our pipeline outperform the previous state-of-the-art method, achieving superior results across both datasets and setting a new benchmark for glomerulus segmentation in WSIs. 
The code and pre-trained models are available at \url{https://github.com/huuquan1994/wsi_glomerulus_seg}.
\end{abstract}
\section{INTRODUCTION}
    \label{sec:intro}
\begin{figure*}[!t]
\centering
\includegraphics[width=0.97\textwidth]{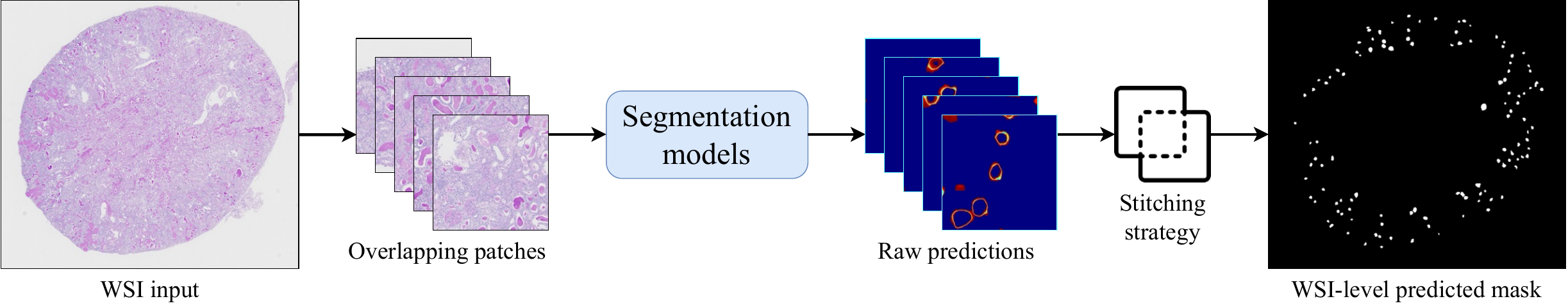}
\caption{
    The overview of our proposed pipeline for WSI glomerulus segmentation. 
}
\label{fig:fig_1}
\end{figure*}
The glomerulus is the main filtering unit of the kidney that helps to retain essential proteins and cells in the blood while filtering out waste and excess substances. 
Glomerular disease is the leading cause of chronic kidney disease (CKD), which is reported to affect over 10\% of the global population, impacting more than 800 million people \cite{kovesdy2022epidemiology}. 
Analyzing and early detection of glomerular disease is essential for diagnosing and treating CKD. 
Previous studies proposed to analyze and detect glomeruli on whole-slide images (WSI) using hand-crafted features or low-level image processing \cite{hirohashi2014automated, kato2015segmental, maree2016approach, govind2018glomerular}. 
However, the complex structure of the glomeruli together with the variability in staining techniques presents significant challenges for traditional image analysis methods. 

Following the tremendous success of deep learning in the field of computer vision, numerous computer-aided diagnostics systems for WSI glomerulus detection have been proposed \cite{bueno2020glomerulosclerosis, jiang2021deep, saikia2023mlp, andreini2024enhancing, tang2024holohisto}. 
Despite the success of the above studies, their results are either reported on the well-known HuBMAP dataset \cite{hubmap2019human} or on private datasets. 
Recently, there are two large and diverse datasets for glomeruli segmentation on kidney tissue that have been released publicly called the mice glomeruli dataset \cite{stritt2020orbit} and the kidney pathology image segmentation challenge dataset (KPIs) \cite{tang2024holohisto}. 
The histopathological image data come from two different species (mice and rats) with several different staining protocols. 
To our knowledge, very few studies have been reported, with only one study available for each dataset \cite{tang2024holohisto, andreini2024enhancing}. 
Thus, more performance analysis on these new and diverse datasets is essential to enrich this field of research. 

In this work, we propose an effective glomeruli segmentation pipeline and comprehensively evaluate its performance on those two datasets. 
Different from \cite{andreini2024enhancing}, which focused on patch-level segmentation, we further extend our evaluation to WSI-level segmentation of the mice glomeruli dataset. 
Additionally, while the work in \cite{tang2024holohisto} used large input patches (i.e., up to 4K) for their end-to-end WSI inference system, we demonstrate that our proposal outperforms theirs using much smaller patches (e.g., $768\times768$). 
To our knowledge, this study is the first to analyze the performance of various segmentation models for both patch-level and WSI-level segmentation tasks on these datasets. 
It is worth noting that with this pipeline, we won 1st place in both tracks of the kidney pathology image segmentation challenge 2024\footnote{\url{https://sites.google.com/view/kpis2024/}}. 
Our contributions are summarized as follows: 
\begin{itemize}
    \item We propose a general glomeruli segmentation pipeline for effectively improving both patch-level and WSI-level segmentation tasks.
    \item We conduct comprehensive performance evaluation and analysis from different segmentation models on the two large and diverse glomeruli datasets. 
    \item Models using our pipeline achieve state-of-the-art results on both datasets. 
    The code and pre-trained models are made publicly available to encourage reproducibility and further research. 
\end{itemize}

\section{METHOD AND MATERIAL}
    \label{sec:mate_method}
    \subsection{The Proposed Segmentation Pipeline}
\begin{figure}[!t]
\centering
\includegraphics[width=0.97\linewidth]{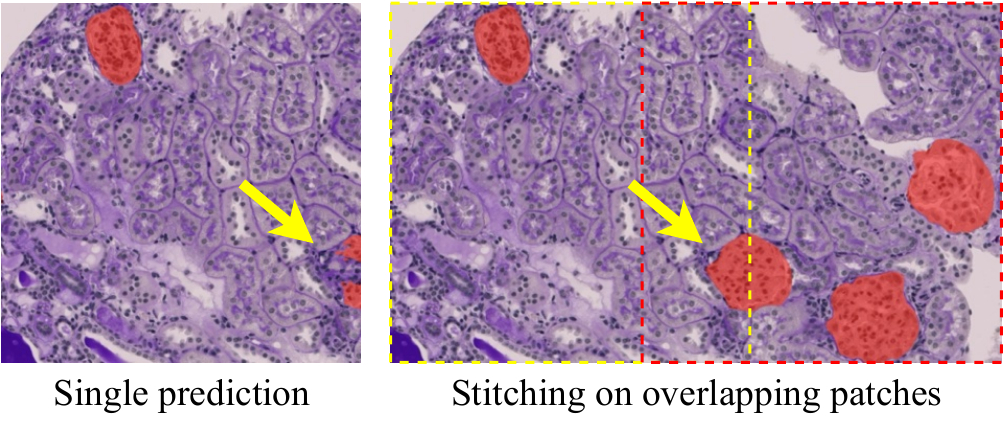}
\caption{
    A comparison between single patch prediction and stitching on overlapping predictions. 
    This strategy effectively increases the detection coverage, especially when glomeruli located near patch image borders. 
}
\label{fig:fig_2}
\end{figure}

Fig. \ref{fig:fig_1} illustrates the overview of our proposed pipeline for WSI glomerulus segmentation. 
Given a WSI input, we first extract overlapping patches by applying a simple sliding window with a stride size. 
After feeding extracted patches to the segmentation models, a stitching strategy is used to combine the raw predictions into a full WSI-level prediction map. 
For overlapping patches, the raw prediction values in the intersecting areas are summed up. 
Finally, the stitched WSI prediction map is normalized using the softmax function. 
This stitching strategy on overlapping patches effectively increases the detection coverage, especially when glomeruli are located near patch image borders. 
Fig. \ref{fig:fig_2} shows a comparison between single patch prediction and stitching on overlapping patches. 
Our stitching strategy on overlapping patches (right, yellow arrow) helps mitigate the problem of losing neighboring context at patch borders (left, yellow arrow). 

For patch-level segmentation, in practice, the patch’s meta information such as the coordinate and its corresponding WSI is usually available (e.g., the top-left coordinate in image filenames), it is straightforward to apply the above stitching strategy to them and crop back to the original patch size based on their coordinates. 

\subsection{Datasets}
In this study, we utilized two datasets, namely the kidney pathology image segmentation (KPIs) challenge dataset \cite{tang2024holohisto} and the mice glomeruli dataset \cite{stritt2020orbit}. 
Table \ref{tab:table_data} shows the statistics of these datasets. 
The specific of each dataset is described as follows: 
\begin{table}[t]
\centering
\caption{Statistics of the mice glomeruli and KPIs datasets}
\label{tab:table_data}
\resizebox{0.97\linewidth}{!}{
\begin{tabular}{llll}
\hline
\multirow{2}{*}{\textbf{Dataset}} & \textbf{Training} & \multicolumn{2}{c}{\textbf{Testing}} \\ \cline{2-4} 
                                  & Patch-level       & Patch-level        & WSI-level       \\ \hline
KPIs \cite{tang2024holohisto}                             & 6,974             & 2,305              & 12              \\
Mice glomeruli \cite{stritt2020orbit}                   & 13,413            & 7,134              & 27              \\ \hline
\end{tabular}
}
\end{table}

\subsubsection{The KPIs Dataset}
This dataset includes 50 high-resolution WSIs of whole mouse kidneys and is released along with the KPIs challenge 2024. 
Tissues were stained with Periodic acid-Schiff (PAS) and captured at $40\times$ digital magnification. 
These WSIs are from four different groups, including three CKD disease models and one normal. 
Specifically, those groups are (1) \textit{Normal} - normal mice; (2) \textit{56Nx} - mice underwent 5/6 nephrectomy; (3) \textit{DN} - eNOS-/-/ lepr(db/db) double-knockout mice; and (4) \textit{NEP25} - transgenic mice that express human CD25 selectively in podocytes. 
The dataset has a total of 9,279 patch images of size $2048\times2048$ with glomerulus annotations extracted from the WSIs. 
Each patch image filename contains the coordinate information (i.e., top-left coordinate) of where it was cropped from its corresponding WSI. 
In addition, there are also 12 WSI ground-truth masks for the WSI-level segmentation evaluation. 

\subsubsection{The Mice Glomeruli Dataset} 
This dataset comprises 88 WSIs with edge sizes ranging from 14,000 to 65,000 pixels. 
The histopathological data come from mice and rats. 
Various staining protocols were used to process including PAS, Hematoxylin (H), Eosin (E), Diaminobenzidine (DAB), Immune Chromogenic Reagent (Fast Red), and three variations of H\&E. 
The dataset is divided into eight WSI groups and has a total of 20,547 glomeruli mask annotations. 
We manually extract patches of size $1024\times1024$ and make sure each patch image contains at least one glomerulus annotation. 
Similar to the KPIs dataset, each patch image filename contains the coordinate information (i.e., top-left coordinate) of where it was cropped from its corresponding WSI. 
Besides patch-level annotations, we also created WSI-level masks to evaluate the WSI-level segmentation task. 
We manually checked and excluded the WSIs that are possibly incomplete in labeling, resulting in 27 WSI-level ground-truth masks. 
More details of this dataset and how to process it can be found in our GitHub repository. 

\section{EXPERIMENTAL RESULTS}
    \label{sec:experiment}
\begin{figure*}[!t]
\centering
\includegraphics[width=0.99\linewidth]{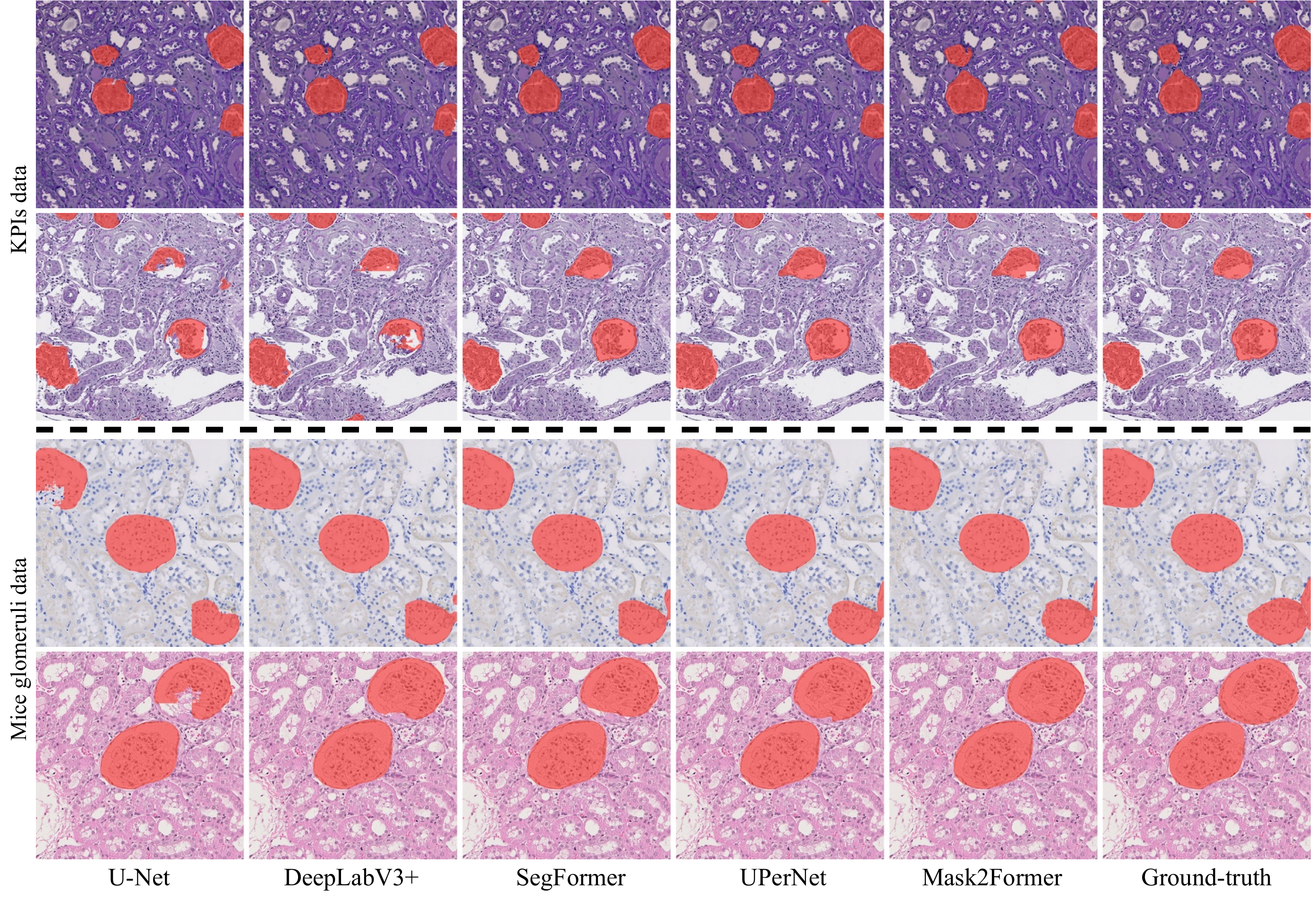}
\caption{
    The visual comparison of patch-level segmentation results on the mice glomeruli dataset (first two rows) and the KPIs dataset (last two rows) from different segmentation models.
}
\label{fig:fig_3}
\end{figure*}
\subsection{Training Segmentation Models}
We built several segmentation models including U-Net \cite{unetronneberger2015}, DeepLabV3+ \cite{chen2018deeplab} with ResNet-50 \cite{he2016deep} backbone, SegFormer-B5 \cite{xie2021segformer}, UPerNet \cite{xiao2018upernet} with ConvNeXt-B \cite{liu2022convnet} backbone, and Mask2Former \cite{cheng2022masked} with Swin-B \cite{liu2021swin} backbone. 
All segmentation models were trained on the patch-level data from two datasets using the same settings. 
Specifically, the models take input of size $768\times768$. 
We used the cross-entropy as the loss function. 
During training, we randomly crop a window size of $768\times768$ together with data augmentation techniques such as random brightness, flips, and blurring. 
The AdamW \cite{loshchilov2017adamw} optimizer was used with the learning rate set to $6\times10^{-5}$. 
We applied the linear rate warm-up for the first 1,500 steps and then polynomial learning rate decay for the rest of the training. 
All models were trained for 30 epochs on 4 GPU NVIDIA V100 with batch size set to 2 per GPU. 
\begin{table}[t]
\centering
\caption{Performance comparison of the patch-level segmentation on the KPIs dataset}
\label{tab:table_1_kpis}
\resizebox{0.98\linewidth}{!}{
\begin{threeparttable}
\begin{tabular}{llllll}
\hline
\textbf{Method} & \textbf{56Nx}  & \textbf{DN}    & \textbf{NEP25} & \textbf{Normal} & \textbf{Avg.}  \\ \hline
U-Net \cite{unetronneberger2015}            & 83.86          & 88.91          & 87.31          & 92.15           & 88.06          \\
\hspace{0.25cm}+ Stitching     & 84.48          & 89.32          & 87.80          & 92.57           & 88.54          \\ \hline
DeepLabV3+ \cite{chen2018deeplab}      & 87.08          & 90.10          & 90.05          & 93.16           & 90.10          \\
\hspace{0.25cm}+ Stitching     & 87.72          & 90.97          & 91.02          & 93.67           & 90.85          \\ \hline
SegFormer \cite{xie2021segformer}      & 93.35          & 94.35          & 93.50          & 94.68           & 93.97          \\
\hspace{0.25cm}+ Stitching     & 93.88          & 94.65          & 93.97          & 94.89           & 94.35          \\ \hline
UPerNet \cite{xiao2018upernet}        & 93.28          & 94.41          & 93.30          & 94.70           & 93.92          \\
\hspace{0.25cm}+ Stitching     & \textbf{93.93} & 94.64          & 93.72          & 94.87           & 94.29          \\ \hline
Mask2Former \cite{cheng2022masked}     & 92.63          & 94.67          & 93.58          & 94.81           & 93.92          \\
\hspace{0.25cm}+ Stitching     & 93.60          & \textbf{95.05} & 93.96          & \textbf{95.07}  & \textbf{94.42} \\ \hline \hline
HoloHisto-4K* \cite{tang2024holohisto}    & 93.77          & 92.45          & \textbf{94.81} & 94.12           & 93.79          \\ \hline
\end{tabular}
\begin{tablenotes}[flushleft]
    \item[*] Result reported from their study
\end{tablenotes}
\end{threeparttable}
}
\end{table}
\begin{table}[t]
\centering
\caption{Performance comparison of the WSI-level segmentation on the KPIs dataset}
\label{tab:table_2_kpis}
\resizebox{0.99\linewidth}{!}{
\begin{threeparttable}
\begin{tabular}{llllll}
\hline
\textbf{Method} & \textbf{56Nx}  & \textbf{DN}    & \textbf{NEP25} & \textbf{Normal} & \textbf{Avg.}  \\ \hline
U-Net \cite{unetronneberger2015}            & 83.60          & 82.51          & 87.61          & 92.75           & 86.62          \\
DeepLabV3+ \cite{chen2018deeplab}      & 89.72          & 91.28          & 91.55          & 93.76           & 91.58          \\
SegFormer \cite{xie2021segformer}       & 94.05          & 94.92          & \textbf{93.97} & \textbf{95.11}  & 94.51          \\
UPerNet \cite{xiao2018upernet}         & 93.95          & 94.61          & 93.41          & 94.94           & 94.23          \\
Mask2Former \cite{cheng2022masked}     & \textbf{94.55} & \textbf{95.23} & 93.79          & 94.99           & \textbf{94.64} \\ \hline \hline
HoloHisto-4K* \cite{tang2024holohisto}    & ---            & ---            & ---            & ---             & 84.54          \\ \hline
\end{tabular}
\begin{tablenotes}[flushleft]
    \item[*] Result reported from their study
\end{tablenotes}
\end{threeparttable}
}
\end{table}
\begin{table*}[t]
\centering
\caption{Performance comparison of the patch-level segmentation on the mice glomeruli dataset}
\label{tab:table_1_orbit}
\resizebox{0.97\linewidth}{!}{
\begin{tabular}{llllllllll}
\hline
\textbf{Method} & \textbf{\begin{tabular}[c]{@{}l@{}}Mouse\\ (Fast Red)\end{tabular}} & \textbf{\begin{tabular}[c]{@{}l@{}}Mouse G1\\ (H\&E)\end{tabular}} & \textbf{\begin{tabular}[c]{@{}l@{}}Mouse G2\\ (H\&E)\end{tabular}} & \textbf{\begin{tabular}[c]{@{}l@{}}Mouse G3\\ (H\&E)\end{tabular}} & \textbf{\begin{tabular}[c]{@{}l@{}}Rat\\ (PAS)\end{tabular}} & \textbf{\begin{tabular}[c]{@{}l@{}}Rat G1\\ (H\&DAB)\end{tabular}} & \textbf{\begin{tabular}[c]{@{}l@{}}Rat G2\\ (H\&DAB)\end{tabular}} & \textbf{\begin{tabular}[c]{@{}l@{}}Rat G3\\ (H\&DAB)\end{tabular}} & \textbf{Avg.}  \\ \hline
U-Net \cite{unetronneberger2015}           & 92.37                                                               & 93.25                                                              & 91.11                                                              & 92.45                                                              & 78.76                                                        & 91.98                                                              & 93.12                                                              & 80.85                                                              & 89.24          \\
\hspace{0.25cm}+ Stitching     & 92.75                                                               & 93.59                                                              & 91.57                                                              & 92.83                                                              & 78.98                                                        & 92.41                                                              & 93.42                                                              & \textbf{81.09}                                                     & 89.58          \\ \hline
DeepLabV3+ \cite{chen2018deeplab}     & 93.00                                                               & 93.85                                                              & 92.39                                                              & 92.49                                                              & 86.98                                                        & 92.97                                                              & 93.54                                                              & 76.35                                                              & 90.20          \\
\hspace{0.25cm}+ Stitching     & 93.21                                                               & 94.06                                                              & 92.75                                                              & 92.72                                                              & 87.58                                                        & 93.29                                                              & 93.73                                                              & 76.57                                                              & 90.49          \\ \hline
SegFormer \cite{xie2021segformer}      & 93.50                                                               & 94.18                                                              & 93.08                                                              & 93.18                                                              & 92.62                                                        & 93.76                                                              & 94.14                                                              & 78.45                                                              & 91.61          \\
\hspace{0.25cm}+ Stitching     & \textbf{93.75}                                                      & 94.42                                                              & 93.30                                                              & 93.33                                                              & 93.16                                                        & 93.95                                                              & 94.27                                                              & 78.50                                                              & 91.84          \\ \hline
UPerNet \cite{xiao2018upernet}        & 93.47                                                               & 94.40                                                              & 93.14                                                              & 93.28                                                              & 92.97                                                        & 93.84                                                              & 94.22                                                              & 79.60                                                              & 91.87          \\
\hspace{0.25cm}+ Stitching     & 93.63                                                               & 94.53                                                              & 93.46                                                              & 93.48                                                              & 93.55                                                        & \textbf{94.01}                                                     & 94.35                                                              & 79.75                                                              & \textbf{92.10} \\ \hline
Mask2Former \cite{cheng2022masked}    & 93.44                                                               & 94.58                                                              & 93.38                                                              & 93.36                                                              & 94.13                                                        & 93.66                                                              & 94.33                                                              & 77.07                                                              & 91.74          \\
\hspace{0.25cm}+ Stitching     & 93.64                                                               & \textbf{94.74}                                                     & \textbf{93.57}                                                     & \textbf{93.70}                                                     & \textbf{94.65}                                               & 93.82                                                              & \textbf{94.42}                                                     & 77.32                                                              & 91.98          \\ \hline
\end{tabular}
}
\end{table*}
\begin{table*}[t]
\centering
\caption{Performance comparison of the WSI-level segmentation on the mice glomeruli dataset}
\label{tab:table_2_orbit}
\resizebox{0.97\linewidth}{!}{
\begin{tabular}{llllllllll}
\hline
\textbf{Method} & \textbf{\begin{tabular}[c]{@{}l@{}}Mouse\\ (Fast Red)\end{tabular}} & \textbf{\begin{tabular}[c]{@{}l@{}}Mouse G1\\ (H\&E)\end{tabular}} & \textbf{\begin{tabular}[c]{@{}l@{}}Mouse G2\\ (H\&E)\end{tabular}} & \textbf{\begin{tabular}[c]{@{}l@{}}Mouse G3\\ (H\&E)\end{tabular}} & \textbf{\begin{tabular}[c]{@{}l@{}}Rat\\ (PAS)\end{tabular}} & \textbf{\begin{tabular}[c]{@{}l@{}}Rat G1\\ (H\&DAB)\end{tabular}} & \textbf{\begin{tabular}[c]{@{}l@{}}Rat G2\\ (H\&DAB)\end{tabular}} & \textbf{\begin{tabular}[c]{@{}l@{}}Rat G3\\ (H\&DAB)\end{tabular}} & \textbf{Avg.}  \\ \hline
U-Net \cite{unetronneberger2015}           & 87.37                                                               & 78.20                                                              & 90.41                                                              & \textbf{75.32}                                                     & 92.03                                                        & 89.51                                                              & 75.97                                                              & 80.43                                                              & 83.66          \\
DeepLabV3+ \cite{chen2018deeplab}     & 89.87                                                               & 65.76                                                              & 86.21                                                              & 66.63                                                              & 89.18                                                        & 90.35                                                              & 67.01                                                              & 87.52                                                              & 80.32          \\
SegFormer \cite{xie2021segformer}      & 83.21                                                               & 76.26                                                              & 86.97                                                              & 68.46                                                              & 84.81                                                        & 89.49                                                              & 61.51                                                              & 86.83                                                              & 79.69          \\
UPerNet \cite{xiao2018upernet}        & 90.73                                                               & \textbf{81.21}                                                     & 92.62                                                              & 73.81                                                              & 92.63                                                        & 92.67                                                              & 78.34                                                              & 91.35                                                              & \textbf{86.67} \\
Mask2Former \cite{cheng2022masked}    & \textbf{90.94}                                                      & 81.18                                                              & \textbf{92.69}                                                     & 68.89                                                              & \textbf{92.93}                                               & \textbf{93.02}                                                     & \textbf{79.44}                                                     & \textbf{92.45}                                                     & 86.44          \\ \hline
\end{tabular}
}
\end{table*}

\subsection{Model Inference}
We evaluate the trained segmentation models under two types of tasks: patch-level and WSI-level segmentation. 
For patch-level data, since all test patch images contain the coordinate information (i.e., top-left position from its corresponding WSI), we perform stitching on patch predictions and then cropping back to the original patch size based on their coordinates. 
For WSI-level data, we simply applied a sliding window with a crop size of $N\times N$ and a stride size of $N/2\times N/2$. 
The stitching strategy is performed on overlapping patches. 
We used $N$ = 2048 and 1024 for the KPIs and mice glomeruli datasets, respectively. 

\subsection{Results and Discussion}
We used Dice score as the evaluation metric. 
The patch-level and WSI-level segmentation results of the mice glomeruli dataset are shown in Table \ref{tab:table_1_kpis} and \ref{tab:table_2_kpis}, while Table \ref{tab:table_1_orbit} and \ref{tab:table_2_orbit} report the results on the KPIs dataset. 
The visual comparison of the patch-level segmentation on the mice glomeruli dataset (first two rows) and the KPIs dataset (last two rows) is shown in Fig. \ref{fig:fig_3}. 
The stitching strategy has proven to be effective as it helped to boost the performance across all models on both datasets. 

On the KPIs dataset, we see the same performance trend on both patch-level and WSI-level segmentation tasks as the lowest performance is from U-Net and DeepLabV3+, followed by SegFormer, UPerNet, and the highest is Mask2Former model. 
We compared our models with HoloHisto-4K \cite{tang2024holohisto}, the previous state-of-the-art method on this KPIs dataset. 
Thanks to our effective stitching strategy, we demonstrated that with a much smaller input size (i.e., $768\times768$) compared to their 4K inputs (i.e., $3840\times2160$), our models achieved superior results, especially the Mask2Former model in the WSI-level segmentation with 94.64 Dice score compared to HoloHisto-4K with 84.54 (over 10 points better, Table \ref{tab:table_2_kpis}). 

On the mice glomeruli dataset, while all models achieved relatively high performances on the patch-level segmentation task (Table \ref{tab:table_1_orbit}), they significantly decreased their Dice score under the WSI-level segmentation task. 
The reason for this is the complex structure of the glomeruli together with the variability in staining techniques on this dataset. 
In addition, since every patch-level image has its ground truth mask, segmentation models might not generalize well on non-annotation patches, thus increasing the false positive rates. 
Notably, DeepLabV3+ and SegFormer models achieved solid performance on patch-level segmentation, however, their Dice scores were considerably lower on the WSI-level segmentation task (decreased by over 12 and 10 points respectively, see Table \ref{tab:table_2_orbit}). 
In this case, introducing some stain color augmentation techniques \cite{tellez2019quantifying} would help to improve the performance. 
Overall, the UPerNet and Mask2Former models have demonstrated robust performance on all benchmarks. 

\section{CONCLUSION}
    \label{sec:conclusion}
    In this work, we propose a pipeline for glomerulus segmentation from whole-slide images of kidney tissue. 
Thanks to the simple yet effective stitching strategy applied to overlapping patches, our pipeline consistently enhances performance across all segmentation models. 
In addition, to our knowledge, this is the first comprehensive evaluation of various segmentation models for both patch-level and WSI-level segmentation tasks on two large and diverse glomeruli datasets. 
By making the pre-trained models publicly available, we believe that our contributions will have a meaningful impact on the field, especially for the early diagnosis of CKD. 
Further improvements to our method are currently being carried out. 



\nocite{*}
\footnotesize{
\bibliographystyle{IEEEtran}
\bibliography{reference}
}

\end{document}